\documentclass[prx, aps, twocolumn, floatfix, superscriptaddress, amsmath,  longbibliography]{revtex4-1}
\usepackage{graphicx}
\usepackage{graphics}
\usepackage{braket}
\usepackage{mathtools}
\usepackage[titletoc]{appendix}
\usepackage{appendix}
\usepackage{float}

\usepackage[dvipsnames]{xcolor}
\definecolor{darkblue}{rgb}{0.0, 0.0, 0.75}

\usepackage{color}
\usepackage[colorlinks=true,
            linkcolor=darkblue,
            urlcolor=darkblue,
            citecolor=darkblue]{hyperref}

\def \bk{{\bf k}}
\def \br{{\bf r}}

\def \mum{\mu \mathrm{m}}

\def \m2D{\mathrm{2D}}
\def \mB{\mathrm{B}}
\def \mms{\mathrm{ms}}
\def \mmms{\mathrm{mm/s}}

\def \mIm{\mathrm{Im}}

\def \mT{\mathrm{T}}

\def \mp{\text{p}}

\def \mHz{\mathrm{Hz}}

\def \cD{\mathcal{D}}

\def \cH{\mathcal{H}}

\def \mT{\mathrm{T}}
\def \mHz{\mathrm{Hz}}

\def \bk{\mathbf{k} }
\def \br{\mathbf{r} }

\def \tA{ {\tilde{A} }}

\def \v0{ {\tilde{v}_0 }}

\def \2F{{_2}F}

\begin{document}
\title{First and second sound in a dilute Bose gas across the BKT transition}
%
\author{Vijay Pal Singh}
\affiliation{Zentrum f\"ur Optische Quantentechnologien and Institut f\"ur Laserphysik, Universit\"at Hamburg, 22761 Hamburg, Germany}
\affiliation{Institut f\"ur Theoretische Physik, Leibniz Universit\"at Hannover, Appelstra{\ss}e 2, 30167 Hannover, Germany}
\author{ Ludwig Mathey}
\affiliation{Zentrum f\"ur Optische Quantentechnologien and Institut f\"ur Laserphysik, Universit\"at Hamburg, 22761 Hamburg, Germany}
\affiliation{The Hamburg Centre for Ultrafast Imaging, Luruper Chaussee 149, Hamburg 22761, Germany}
\date{\today}
%
%
\begin{abstract}
We study the propagation of the two sound modes in two-dimensional Bose gases across the Berezinksii-Kosterlitz-Thouless (BKT) transition using classical-field dynamics, which is motivated by recent measurements of Christodoulou {\it et al.} Nature \textbf{594}, 191 (2021). 
Based on the dynamic structure factor, we identify the two sound modes as the Bogoliubov (B) and the non-Bogoliubov (NB) sound mode below the transition, and as the diffusive and the normal sound mode above the transition. 
The NB sound mode velocity is higher than the B sound mode velocity, which we refer to as the weak-coupling regime of the sound modes. 
We excite the sound modes by driving the system as in the experiment and by perturbing the density with a step-pulse perturbation, as a secondary comparison.
The driven response depends on the driving strength and results in higher velocities for the B sound mode at high temperatures near the transition, compared to the sound results of the dynamic structure factor and step-pulse excitation.   
We show that the higher mode velocity has a weak temperature dependence across the transition, which is consistent with the experimental observation.  
\end{abstract}

\maketitle
%
%

\section{Introduction}\label{sec:intro}
Superfluidity in quantum fluids  is in general accompanied by  the phenomenon of two sound modes, namely, first and second sound, 
which is supported by Landau-Tisza's two-fluid hydrodynamic theory \cite{Tisza1938, Landau1941}.
The intriguing phenomenon of second sound was first observed in liquid helium and was described as an entropy wave based on the two-fluid hydrodynamic theory \cite{Peshkov}. Ultracold atoms expanded the scope of this study by a wide range of trappable quantum liquids including features such as reduced dimensionality and tunable interactions. Two sound modes were measured in a three-dimensional (3D) unitary Fermi gas \cite{Sidorenkov2013}, the BEC-BCS crossover \cite{Hoffmann2021}, and dilute 2D and 3D Bose gases \cite{Hadzibabic2021, Hilker2021}.

   Contrary to liquid helium, ultracold gases form a wide range of weakly interacting quantum fluids, which undergo an intriguing interplay of sound modes between hydrodynamic and non-hydrodynamic regimes \cite{Stamper1998, SinghSF, Hilker2021}.  
Based on the hydrodynamic theory at zero temperature,  the second-sound velocity ($v_2=v_1/\sqrt{D}$) is either below the first-sound velocity $v_1$  for $D=3$ and $2$ dimensions or equal to  $v_1$  for $D=1$ dimension \cite{Matveev2018}. 
Hydrodynamic theory does not support the sound velocity being higher than the Bogoliubov velocity. 
However, such regimes at low temperatures were predicted in dilute 3D and 2D Bose gases in the weak-coupling regime  \cite{Ilias, SinghSS}.    
  
 Sound propagation in 2D quantum fluids is of particular interest because the superfluid density undergoes a universal jump of $4/\lambda^2$ at the Berezinksii-Kosterlitz-Thouless (BKT) transition \cite{Berezinski1972, Kosterlitz1973, Minnhagen1987}, where $\lambda$ is the thermal de Broglie wave length.  
This has attracted interest to study sound propagation in ultracold 2D quantum gases both experimentally \cite{Dalibard2018, Bohlen2020, Hadzibabic2021} and theoretically \cite{Ozawa2014, Liu2014, OtaTwo, Ota2018,  Salasnich2018, SinghSS, Zhigang2020, Krzysztof, Salasnich2020, Furutani2021}.
In Ref.  \cite{SinghSS}  we discussed and gave numerical evidence for the weak and strong coupling regimes of the sound modes. 
For the weak-coupling regime, we found that the non-Bogoliubov (NB) sound mode has higher velocity than the Bogoliubov (B) sound mode. We referred to this as a non-hydrodynamic regime. 
For the strong-coupling regime, we showed that the B sound mode velocity is higher than the NB sound mode velocity. 
This was consistent with a hydrodynamic scenario. 
Furthermore, we found that the two sound modes undergo a temperature-dependent hybridization between these two coupling regimes \cite{SinghSF}. We note that  for a finite-size system the BKT transition manifests itself as a crossover \cite{Pilati2008, Matt2010, sf_2019,  SinghJJ},  rather than a sudden jump.

     Recently, Ref. \cite{Hadzibabic2021} reported the measurement of the two sound modes in a homogeneous 2D Bose gas of $^{39}$K atoms across the BKT transition. 
The density response of the driven system is measured as a function of the driving frequency, allowing the detection of both sound modes. 
The velocity of the lower sound mode decreases with increasing temperature and vanishes above the transition temperature $T_c$, 
whereas the velocity of the higher sound mode is higher than the Bogoliubov velocity at zero temperature and displays a weak temperature dependence across $T_c$. 
The measurements are compared to the two-fluid theory in an infinite system \cite{OtaTwo}, which predicts a jump in the velocities at $T_c$ and does not describe the temperature dependence of the higher sound mode at $T/T_c<0.75$.

  In this paper, we use classical-field simulations to study the propagation of the sound modes in 2D Bose gases for the experimental parameters of Ref. \cite{Hadzibabic2021}. 
The dynamic structure factor (DSF) of the unperturbed cloud shows the Bogoliubov (B) and non-Bogoliubov (NB) sound modes below the transition temperature and the diffusive and normal sound modes above the transition temperature.  
This allows us to determine the two sound velocities across the BKT transition, independent of an external probe,  serving as a benchmark for the results of the density probes. 
We implement the experimental method to excite the sound modes by driving the system \cite{Hadzibabic2021}. 
We find a driving-strength dependent density response and the excitation of two well-resolved sound peaks is observed for strong driving strengths only. 
As a secondary comparison, we excite the two sound modes using a step-pulse perturbation of the density \cite{SinghSS, Hoffmann2021}. 
Finally, we compare the sound velocity results of the driven response, the step-pulse perturbation and the DSF with the measurements   \cite{Hadzibabic2021}. The measured higher-mode velocity within the experimental error agrees with the simulation results at all temperatures, even at  temperatures where the hydrodynamic prediction fails. 
The measured lower velocity shows a shift to higher velocities compared to the results of the DSF and step-pulse perturbation, which is captured by the simulation results of the driven response at high temperatures. 
This increase due to the nonlinear response of the strong driving partially captures the experimental observations.

This paper is organized as follows. 
In Sec. \ref{sec:method} we describe the simulation method and the excitation protocols. 
In Sec. \ref{sec:dsf} we calculate the dynamic structure factor to characterize the sound modes.
In Sec. \ref{sec:probes} we present the results of the driven response and the step-pulse perturbation.
In Sec. \ref{sec:comp} we compare the simulation and the measurements of the two sound velocities. 
We conclude in Sec. \ref{sec:conclusion}.

\begin{figure*}[]
\includegraphics[width=1.0\linewidth]{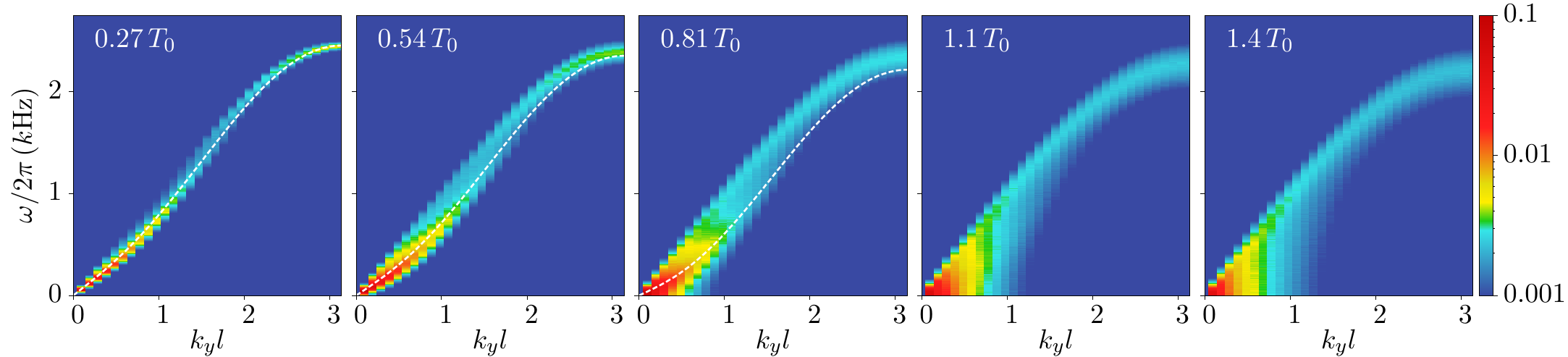}
\caption{Excitation spectra of a 2D Bose gas for the density $n=3 \, \mum^{-2}$ and the interaction $\tilde{g}=0.64$, which are the same as the experimental values in Ref.  \cite{Hadzibabic2021}. 
Dynamic structure factor $S(\bk, \omega)$ as a function of the wave vector $\bk= k_y$ and the frequency $\omega$ is shown for various $T/T_0$ across the BKT transition, where $T_0$ is the estimate of the transition temperature. 
The white dashed line is the Bogoliubov dispersion determined using the numerically obtained superfluid density $n_s(T)$; see text. 
   }
\label{Fig:dsf}
\end{figure*}

\section{System and methodology}\label{sec:method}
We simulate a bosonic cloud of $^{39}$K atoms confined to 2D motion in a box potential. 
This geometry  was  used in Ref. \cite{Hadzibabic2021}. 
The system is described by the Hamiltonian 
\begin{equation} \label{eq_hamil}
\hat{H}_{0} = \int d \br \, \Big[\frac{\hbar^2}{2m} \nabla \hat{\psi}^\dagger({\bf r})  \cdot \nabla \hat{\psi}({\bf r}) 
 + \frac{g}{2} \hat{\psi}^\dagger({\bf r})\hat{\psi}^\dagger({\bf r})\hat{\psi}({\bf r})\hat{\psi}({\bf r})\Big].
\end{equation}
$\hat{\psi}$ ($\hat{\psi}^\dagger$) is the bosonic annihilation (creation) operator.
$g = \tilde{g}\hbar^2/m$ is the 2D interaction parameter, with $\tilde{g}= \sqrt{8 \pi} a_s/\ell_z$ being the dimensionless interaction and 
$m$ the atomic mass.
$a_s$ is the 3D scattering length and $\ell_z$ is the harmonic-oscillator length of the confining potential in the transverse direction.
We use the same density $n=3\, \mum^{-2}$ and the same $\tilde{g}=0.64$,  as in the experiments \cite{Hadzibabic2021}.
We choose a square box of size $L_x \times L_y = 32 \times 32 \, \mum^2$ and various temperatures $T/T_0$. 
We use the temperature $T_0 = 2\pi n \hbar^2 /(m k_\mB \cD_c)$, with the critical phase-space density $\cD_c= \ln(380/\tilde{g})$, as the temperature scale, see Refs. \cite{Prokofev2001, Prokofev2002}. 
This scale gives an estimate of the critical temperature $T_c$ of the BKT transition.
For the simulations we discretize space on a lattice of size $N_x \times N_y = 64 \times 64$ and a discretization length $l=0.5\, \mum$.
We note that $l$ is chosen to be smaller than the healing length $\xi= \hbar/\sqrt{2mgn}$ and the thermal de Broglie wave length \cite{Castin2003}.
We use the classical-field method of Refs. \cite{Singh2017, SinghJJ}.
According to this method, we replace the operators $\hat{\psi}$ in Eq. \ref{eq_hamil} and in the equations of motion by complex numbers $\psi$.
We sample the initial states from a grand-canonical ensemble of a chemical potential $\mu$ and a temperature $T$ via a classical Metropolis algorithm. We propagate the state using the equations of motion to obtain the many-body dynamics.
To excite the sound modes we add the perturbation term 
 \begin{align}
 \cH_\mp = \int d \br V(\br, t) n(\br, t),
 \end{align}
where $V(\br, t)$ is the perturbation potential that couples to the density $n(\br, t)$ at the location $\br = (x, y)$ and time $t$.
Within linear response theory, the induced density fluctuation  $\delta n(\bk, \omega)$ is described in terms of the density response function $\chi_{nn} (\bk, \omega)= \delta n(\bk, \omega)/V(\bk, \omega)$,  where $V(\bk, \omega)$ is the Fourier transform of $V(\br, t)$. 
This allows us to determine the collective modes of the system.

\begin{figure*}[]
\includegraphics[width=1.0\linewidth]{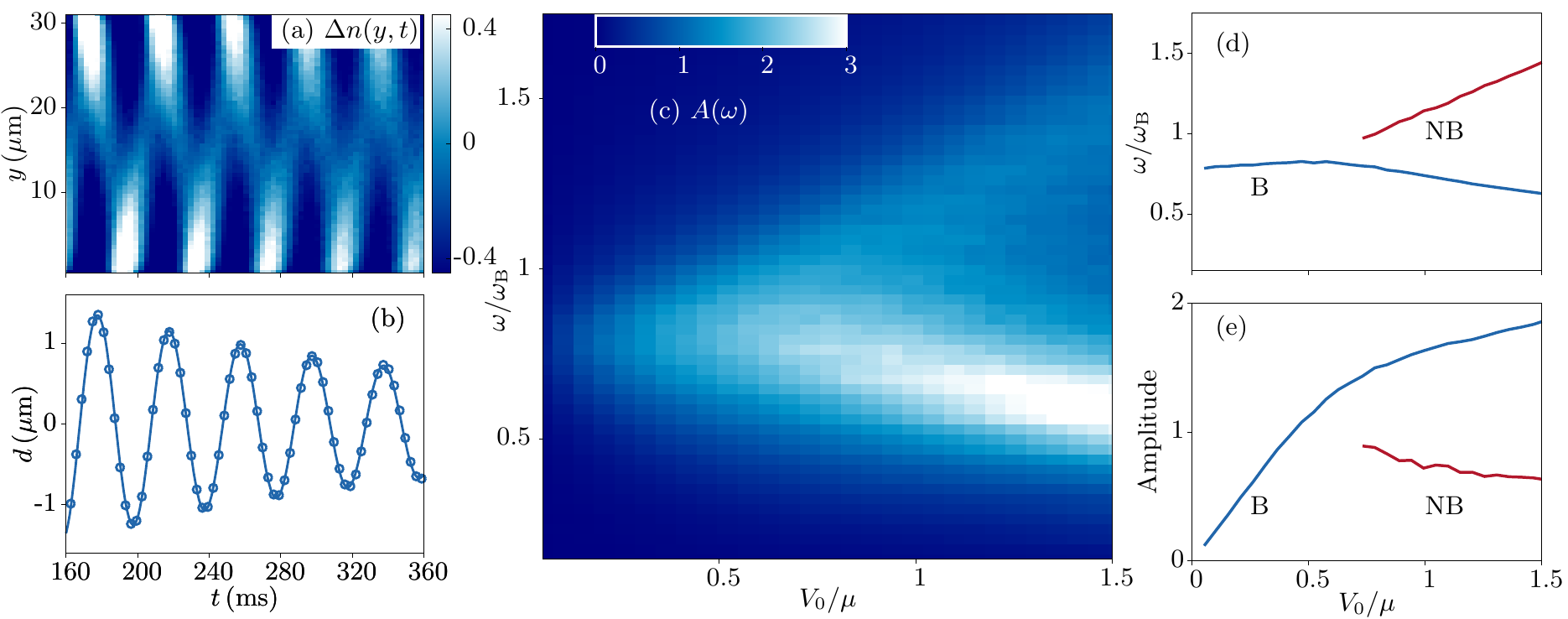}
\caption{Detecting both sound modes via periodic driving of the center of mass of the cloud. 
(a) Time evolution of the density profile $\Delta  n(y, t)$, averaged over the $x$ direction and the ensemble, for $V_0/\mu=0.8$,  $\omega/\omega_\mB=0.74$ and $T/T_0 = 0.54$. 
(b) Displacement $d(t)$ of the cloud's center of mass and the fit $f(t) = [R \sin(\omega t) - A  \cos(\omega t)] \exp(-\kappa t)$ (continuous line) using the fitting parameters $A$, $R$ and $\kappa$. 
(c) $A(\omega)$ response as a function of $V_0/\mu$  and $\omega/\omega_\mB$.
(d, e) show the determined values of the mode frequency and the amplitude; see text.  
B and NB denote the Bogoliubov and non-Bogoliubov sound mode, respectively.
   }
\label{Fig:shaking}
\end{figure*}

 We first implement the experimental method of exciting both sound modes, as in Ref. \cite{Hadzibabic2021}.
We drive the system along the $y$ direction using $V(\br, t)= V_0  \sin(\omega t) \times (y-L_y/2)/L_y$, 
where $V_0$ is the driving strength and $\omega$ is the driving frequency.
This predominantly excites the longest wavelength sound modes with the wavevector $k_L=\pi/L_y$ \cite{Navon2016}.
For each $\omega$, we calculate the time evolution of the density profile $\Delta n(y, t)=n(y, t) - n$, 
which is averaged over the ensemble and the $x$ direction. 
The driving protocol results in center-of-mass oscillations of the cloud, as shown in Fig. \ref{Fig:shaking}(a). 
Assuming that the density fluctuation corresponds to the lowest excitation mode in the box, we fit $\Delta n(y, t)$ with the function $n(y, t) = b_0 + b(t) \sin[\pi (y-L_y/2)/L_y]$ using $b_0$ and $b(t)$ as the fitting parameters.
From $b(t)$, we obtain the center-of-mass displacement $d(t)=(1/L_y) \int_0^{L_y}  dy\, y n(y, t)  = 2b(t)L_y/\pi^2$.
To calculate the steady-state response, we choose the time evolution between $160-360\, \mms$, which is fixed for every $\omega$.
Fitting $d(t)$ with the function $f(t) = [ R(\omega) \sin(\omega t) - A(\omega) \cos(\omega t)]  \exp(-\kappa t)$  
enables us to determine the reactive $(R)$ and absorptive $(A)$ response, 
where we have included an additional fit parameter $\kappa$ as global damping rate. 
Using the Fourier decomposition $V(k_L, \omega)= 4 V_0/\pi^2$, the $A(\omega)$ response
yields $\mIm \chi_{nn} (k_L, \omega)  = n \pi^4 A(\omega)/(8V_0L_y)$ and thus allows to 
determine the dynamic structure factor $S(k_L, \omega)$ \cite{Hadzibabic2021}.

As second method, we employ a step-pulse perturbation of the density, which is motivated by Refs. \cite{SinghSS, Hoffmann2021}. 
To perturb the density  we use the Gaussian potential $V(\br, t)= V_0(t) \exp[- (y-y_0)^2/(2 \sigma^2) ]$, which is centered at $y_0= L_y/2$. 
$V_0(t)$ is the time-dependent strength and $\sigma$ is the width. 
We suddenly turn on and off this potential for a short perturbation time of about $0.5\, \mms$, 
which excites sound pulses as shown in Fig. \ref{Fig:step}.

 \begin{figure}[]
\includegraphics[width=1.0\linewidth]{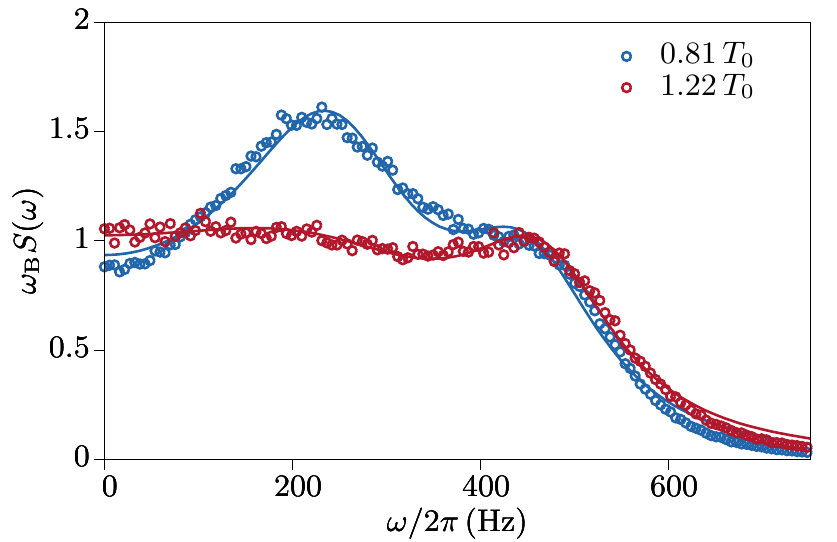}
\caption{$S(k, \omega)$ plots at $k/k_\xi= 0.35$ for $T/T_0=0.81$ (blue circles) and $1.22$ (red circles).
$k_\xi$ is the wave vector below which the Bogoliubov dispersion has a linear momentum dependence. 
The continuous lines are the fits with the two-mode dynamic structure factor in Eq. \ref{eq:fit}.
}
\label{Fig:modes}
\end{figure}

\section{Dynamic structure factor}\label{sec:dsf}
To characterize the sound modes we calculate the dynamic structure factor (DSF)
\begin{align}
S(\bk, \omega) =  \langle |n(\bk, \omega)|^2  \rangle,
\end{align}
where $n(\bk, \omega)$ is the Fourier transform of the density $n(\br, t)$ in space and time. We define $n(\bk, \omega)$ as 
\begin{align}
n(\bk, \omega) = \frac{1}{\sqrt{N_l T_s}} \sum_i \int dt \, e^{-i(\bk \br_i - \omega t)} n(\br_i, t).
\end{align}
$N_l= N_x N_y$ is the number of lattice sites and $T_s = 160 \, \mms$ is the sampling time for the numerical Fourier transform. 
The DSF gives the overlap of the density degree of freedom with the collective modes. 
In Fig. \ref{Fig:dsf} we show $S(\bk, \omega)$ as a function of the wave vector $\bk=k_y$ and the frequency $\omega$ for various $T/T_0$. 
At low temperatures it primarily shows one excitation branch, while at intermediate temperatures it shows two excitation branches.
Above the transition temperature, it displays the diffusive mode at low momenta and the excitation branch of the normal sound mode in a thermal gas.  
We compare these spectra with the Bogoliubov dispersion $\hbar \omega_k = \sqrt{\epsilon_k(\epsilon_k + 2 g n_s )}$, 
where $\epsilon_k = 2J[1- \cos (k_y l)]$ is the free-particle dispersion on the lattice introduced for simulations and $J= \hbar^2/(2ml^2)$ is the tunneling energy.  
We numerically determine the superfluid density $n_s(T)$ from the current-current correlations; see Ref. \cite{SinghSS}.
We show the Bogoliubov prediction in Fig. \ref{Fig:dsf}, 
which agrees with the lower branch at all $k$ for all $T/T_0$ below the transition temperature.
This enables us to identify the lower branch as the Bogoliubov (B) mode and the higher branch as the non-Bogoliubov (NB) mode. 
At low temperatures the spectral weight is on the B mode, while at intermediate temperatures both B and NB modes are visible. 
The broadening of the B mode increases with increasing temperature, which occurs due to Landau damping \cite{Chung2009}.  
Above the transition temperature,  the B mode transforms into the diffusive mode and the NB mode continuously connects to the normal sound mode.

%

In experiments, the DSF is measured via the density response $\chi_{nn} (\bk, \omega)$ that describes the density fluctuation created by a perturbing potential. Thus, the density response is a useful tool to identify the sound modes using density probes, as we show in Sec. \ref{sec:probes}. 
For $k_\mB T \gg \hbar \omega$, the density response relates the DSF via 
$S(\bk, \omega)= -  k_\mB T \mIm \chi_{nn} (\bk, \omega)/(\pi n \omega)$ \cite{Hohenberg1964, Griffin2009}.
The two sound modes are supported by Landau-Tisza's two-fluid hydrodynamic theory, yielding the density response \cite{Griffin2009}
\begin{align}\label{eq:nn}
\chi_{nn} (\bk, \omega) = \frac{n k^2}{m} \frac{\omega^2-v^2 k^2}{ (\omega^2-v_1^2 k^2) (\omega^2-v_2^2 k^2)  },
\end{align}
which has poles at the velocities $v_1$ and $v_2$ corresponding to the two sound modes.  
$v^2 =  T s^2 n_s/(c_v n_n)$ denotes an additional velocity, where $s$ is the entropy, $c_v$ is the heat capacity, $n_s$ is the superfluid density, and $n_n$ is the normal fluid density. 
Following Eq. \ref{eq:nn} and including linear damping \cite{Hohenberg1965}, we fit the simulated $S(\bk, \omega)$ with $S(\omega) =S_1(\omega)+S_2(\omega)$ for each $\bk=k_y$, where 
\begin{align}\label{eq:fit}
S_{1,2} (\omega) = \frac{x_{1,2} \omega_{1,2}^2 \Gamma_{1,2}} { (\omega^2 - \omega^2_{1,2} )^2  + ( \omega \Gamma_{1,2})^2 }.
\end{align}
The amplitudes $x_{1,2}$, mode frequencies $\omega_{1,2}$ and damping rates $\Gamma_{1,2}$ are the fit parameters. 
As an example, in Fig. \ref{Fig:modes} we show the simulated $S(\bk, \omega)$ at $k/k_\xi=0.35$ for $T/T_0=0.81$ and $1.22$. 
The wave vector $k_\xi \equiv \sqrt{2}/\xi = 2.8\, \mum^{-1}$ sets the momentum scale, below which the Bogoliubov dispersion has a linear momentum dependence. We fit these spectra with Eq. \ref{eq:fit} and show their fits as the continuous lines in Fig. \ref{Fig:modes}. 
The two-mode feature of the DSF is captured by Eq. \ref{eq:fit}.
At $T/T_0=0.81$, the lower (B) sound peak has higher spectral weight than the higher (NB) sound peak. 
Above the transition temperature at $T/T_0=1.22$, the B mode vanishes and results in the diffusive mode, 
while the NB mode becomes the normal sound mode. 
This numerical observation of the two sound peaks is consistent with the measured spectra \cite{Hadzibabic2021}, which we discuss below. 
To determine the sound velocities, we perform fits for various $k$ below $k_\xi$ and determine the velocities from the linear slope of $\omega_{1,2}/k$. 
When there is mainly one dominant mode at low and high temperatures,
we fit with a single function in Eq. \ref{eq:fit}. 
The results of the two sound velocities are presented in Fig. \ref{Fig:vel}.

\section{Excitation of density pulses}\label{sec:probes}
The two sound modes that we find in the dynamic structure factor can be measured by exciting density pulses \cite{Sidorenkov2013, Hoffmann2021, Hadzibabic2021}. In the following, we present the method of periodic modulation \cite{Hadzibabic2021} and a potential quench of the local density  \cite{Hoffmann2021, SinghSS}.

\begin{figure}[t]
\includegraphics[width=1.0\linewidth]{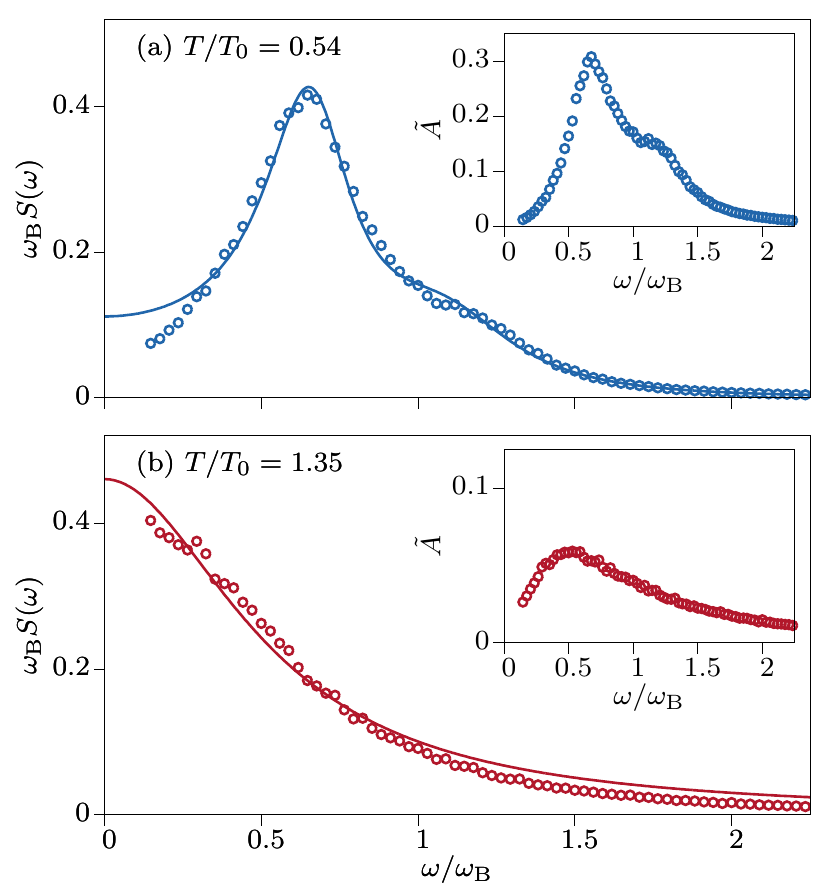}
\caption{Dimensionless $\tilde{A}(\omega)$ response (inset) and the corresponding dynamic structure factor $S(\omega)$ are shown at $T/T_0= 0.54$ (a) and $1.35$ (b). We used $V_0/\mu = 1$.
The continuous line in (a) is the fit with Eq. \ref{eq:fit}, 
whereas the continuous line in (b) is the Lorentzian fit centered at $\omega=0$. 
}
\label{Fig:adsf}
\end{figure}

\subsection{Periodic driving}\label{sec:excitation}
 We first present the method of exciting both sound modes via periodic driving of the center of mass of the cloud, 
 as described in \cite{Hadzibabic2021} and Sec. \ref{sec:method}. 
The driving potential is directed along the $y$ direction and sinusoidally oscillates in time at frequency $\omega$.  
In Fig. \ref{Fig:shaking}(a) we show the resulting time evolution of the density profile $\Delta n(y, t)$ for      
$V_0/\mu=0.8$,  $\omega/\omega_\mB=0.74$ and $T/T_0 = 0.54$. 
$\mu = gn$ is the mean-field energy and $\omega_\mB = v_\mB k_L $ is the Bogoliubov frequency, 
which results in $\omega_\mB/(2\pi) = 34 \, \mHz$ for $v_\mB= \sqrt{gn/m}=2.25\, \mmms$. 
The driving protocol excites  center-of-mass oscillations of the cloud at $\omega/\omega_\mB$, 
from which we determine the displacement $d(t)= 2b(t)L_y/\pi^2$; see Sec. \ref{sec:method}.
In Fig. \ref{Fig:shaking}(b) we show $d(t)$ and the corresponding fit with the function $f(t) = [R(\omega) \sin(\omega t) - A(\omega) \cos(\omega t)] \exp(-\kappa t)$, which allows us to determine the reactive $(R)$ and absorptive $(A)$ response.   
We find that the damping of the oscillations, determined by $\kappa$, depends on $\omega$ and $V_0$ of the driving potential. 
In Fig. \ref{Fig:shaking}(c) we show the $A(\omega)$ response determined as a function of $V_0/\mu$ and $\omega/\omega_\mB$.
For low $V_0/\mu$, $A(\omega)$ primarily displays one maximum that corresponds to the B mode. 
The location of this maximum occurs below the zero-temperature prediction $\omega/\omega_\mB = 1$, 
which is due to the thermal broadening of the phonon modes at nonzero temperatures \cite{SinghSS}. 
For higher $V_0/\mu$ near and above $1$, $A(\omega)$ shows two maxima corresponding to the B and NB modes. 
Interestingly, the separation between the peak locations in frequency space increases with increasing $V_0/\mu$, which is due to the nonlinear response of the system, that sets in for larger perturbations beyond the linear response, as we describe below.

   To determine the peak amplitude and the frequency, we fit $A(\omega)$ with the function $A_{1, 2} (\omega) = \omega S_{1, 2} (\omega)$, based on the relation $S(\bk, \omega) \propto A(\omega)/\omega$, see Sec. \ref{sec:dsf} and Eq. \ref{eq:fit}.
In Figs. \ref{Fig:shaking}(d) and (e) we plot the frequency and the amplitude of each sound peak, determined via individual fitting, as a function of $V_0/\mu$. 
For low driving strengths up to $V_0/\mu \sim 0.6$, the B-mode frequency remains qualitatively unchanged and is about $\omega_1/\omega_\mB \approx 0.8$. In this weak perturbation regime, the B-mode amplitude increases linearly, which is a characteristic of linear response. 
At higher $V_0/\mu$, the B-mode amplitude shows nonlinear behavior, where its frequency decreases and drops to 
$\omega_1/\omega_\mB = 0.62$ for $V_0/\mu = 1.5$.
 The higher (NB) sound peak is resolved only above $V_0/\mu \gtrsim 0.7$. Contrary to the B-mode, the NB-mode frequency increases with increasing $V_0/\mu$, while its amplitude decreases. 
This reduction of the B-mode frequency and enhancement of the NB-mode frequency derives from the decreasing superfluid and increasing normal fluid density due to stronger probing, respectively.  
To  determine the sound velocities $v_{1,2}=\omega_{1,2}/k_L$, we use the $A(\omega)$ response in nonlinear regime at $V_0/\mu=1$, 
where the two sound peaks are well resolved and motivated by the probing regime of Ref. \cite{Hadzibabic2021}. 
We note that at this regime the frequencies of the modes are weakly renormalized by the probe, compared to the linear response regime. We show the results of $v_{1,2}$ for various $T/T_0$ in Fig. \ref{Fig:vel}.

\begin{figure*}[t]
\includegraphics[width=1.0\linewidth]{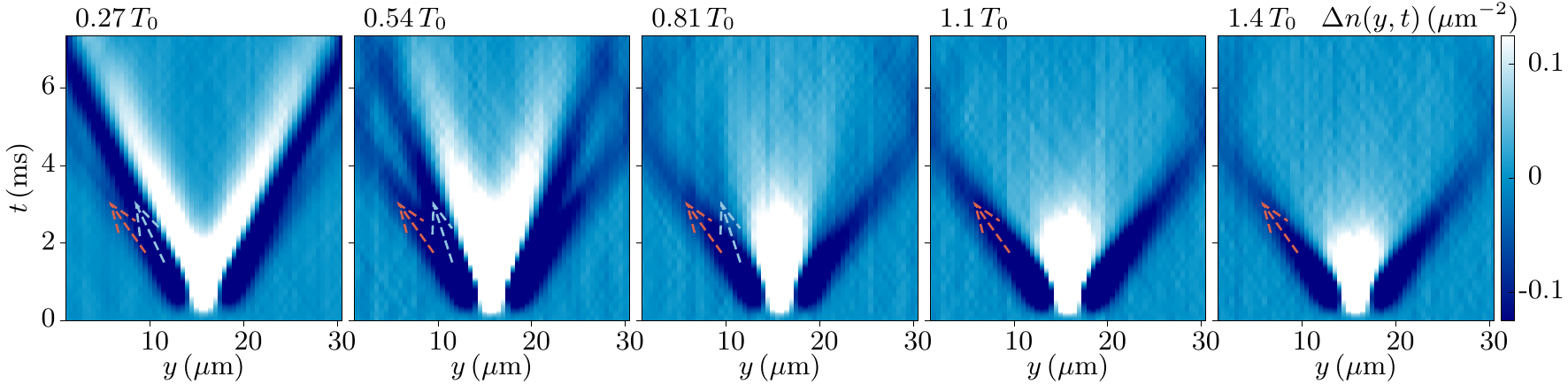}
\caption{Excitation of two sound pulses via a step-pulse perturbation of the local density.
Time evolution of the density $\Delta  n(y, t)$ displays the propagation of two sound modes below the transition and one sound mode  above the transition. The blue (red) arrow denotes the propagation of the slow (fast) sound mode. 
The attractive potential produces a density increase at the location of the perturbation, 
which results in an additional excitation (white pulses) after the potential is switched off. 
  }
\label{Fig:step}
\end{figure*}

 We now relate the $A(\omega)$ response to the dynamic structure factor (DSF) both below and above the transition temperature. 
 For this, we calculate the $A(\omega)$ response using $V_0/\mu = 1$ at $T/T_0=0.54$ and $1.35$. 
 In Fig. \ref{Fig:adsf}, we show the dimensionless response $\tA = \pi^3 m v_\mB^2 A/(8V_0 L_y)$ and the corresponding DSF
 $S = k_\mB T \tA/(m v_\mB^2 \omega)$ below and above the transition temperature. 
For $T/T_0=0.54$, $\tA(\omega)$ shows the two sound peaks, which are also visible in the corresponding $S(\omega)$. 
The two-peak structure is captured by the DSF in Eq. \ref{eq:fit} and is consistent with the simulated DSF of unperturbed cloud in Fig. \ref{Fig:dsf}. 
Above the transition temperature, $\tA(\omega)$  primarily shows the diffusive sound peak and the higher mode is not discernible. 
This absence of the higher sound peak is in contrast with the measured response \cite{Hadzibabic2021} as well as the simulated DSF in Fig. \ref{Fig:dsf} and the step-pulse excitation in Fig. \ref{Fig:step}, which show both diffusive and higher sound modes. 
The reason for this discrepancy can be the decay of the oscillations shown in Fig. \ref{Fig:shaking}(b), 
which is not discernible in the measurements  \cite{Hadzibabic2021}.
In Fig. \ref{Fig:adsf}(b) we show $S(\omega)$ corresponding to $\tA(\omega)$ at $T/T_0= 1.35$, 
which results in the diffusive sound mode centered at $\omega=0$. 
We fit  $S(\omega)$ with the Lorentzian $f(\omega)= x_\mT \Gamma_\mT/(\omega^2 + \Gamma_\mT^2)$ 
using $x_\mT$ and  $\Gamma_\mT$ as the fitting parameters.
The width of the diffusive mode yields the thermal diffusivity $D_\mathrm{T}= \Gamma_\mT/k_L^2 = (7.2 \pm 0.2) \hbar/m$, 
which agrees with the measured $D_\mathrm{T} = (5 \pm 2) \hbar/m$ \cite{Hadzibabic2021}.
Furthermore, our result of $D_\mathrm{T}$ is above the sound diffusivities measured in strongly interacting 2D Fermi gases \cite{Bohlen2020}.

\subsection{Step-pulse perturbation}\label{sec:step}

\begin{figure}[]
\includegraphics[width=1.0\linewidth]{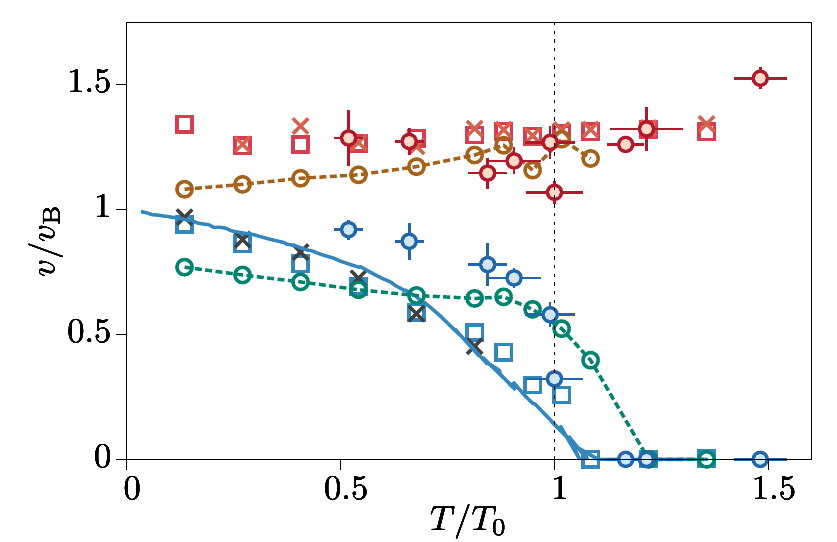}
\caption{Temperature dependence of the two sound mode velocities across the BKT transition. 
Results of the dynamic structure factor (blue and red squares), the driven response with $V_0/\mu=1$ (open circles connected with a dashed line)  and the step-pulse perturbation (black and red crosses) are compared with the measurements of the two sound velocities (blue and red filled circles) of Ref. \cite{Hadzibabic2021}.  
The continuous line is the Bogoliubov estimate $v_{\mB, T}$ based on the numerically determined superfluid density; see text.
}
\label{Fig:vel}
\end{figure}

In this section we demonstrate the method of  step-pulse perturbation, which excites sound modes by locally perturbing the density. 
The perturbation sequence is described in Sec.  \ref{sec:method} and the perturbation potential is a Gaussian, 
for which we use the strength $|V_0|/\mu$ in the range $0.2-1.0$ and the width $\sigma/\xi=2.9$, 
where $\xi=0.51\, \mum$ is the healing length. 
In Fig. \ref{Fig:step} we show the time evolution of the density profile $\Delta n(y, t)$. 
The perturbation potential was turned on for about $0.5\, \mms$, which excites two (fast and slow) sound pulses below the transition temperature and one sound pulse above the transition temperature.
The increased density at the location of the perturbation results in an additional pulse after the perturbation is turned off.
Below $T/T_0  \sim 1$, the fast and slow pulse initially overlap and then eventually separate into two pulses propagating at different velocities, 
as shown in Fig. \ref{Fig:step}. 
The fast pulse corresponds to the NB mode, whereas the slow pulse represents the B mode. 
In the long-time evolution the fast pulse bounces off the wall of the box and continues propagating towards the center.   
At low temperatures, the amplitude of the B mode is higher than the NB mode, 
whereas at higher temperature $T/T_0=0.81$, the NB mode has higher amplitude than the B mode. 
This is consistent with the spectral weights of the modes in the dynamic structure factor in Fig. \ref{Fig:dsf}. 
Near the transition temperature $T/T_0 \sim 1$, the propagation of the B mode vanishes and results in the diffusive sound mode.  
Above the transition temperature, the time evolution primarily shows the propagation of the normal sound mode and also diffusive dynamics at the location of the perturbation.
To obtain the sound velocities, we fit the density profile with one or two Gaussians to determine the locations of one or two density pulses. 
The time dependence of the locations gives the sound velocities, which we show in Fig. \ref{Fig:vel}.

\section{Comparison to experiments}\label{sec:comp}

  In Fig. \ref{Fig:vel}, we combine our simulation results of the sound velocities and compare them with the measurements \cite{Hadzibabic2021}.
The temperature dependence of the two mode velocities determined from the dynamic structure factor (DSF) of unperturbed cloud serves as a benchmark for the results based on the density response involving external perturbations.
The higher-mode velocity displays a weak temperature dependence at all temperatures and no signature of a jump at the transition. 
On the other hand, the lower-mode velocity decreases with increasing temperature and vanishes above the transition temperature. 
We compare this result with the nonzero temperature Bogoliubov estimate $v_{\mB, T}=\sqrt{g n_s(T)/m}$, which is obtained using the numerically determined superfluid density $n_s(T)$; see, for details, Ref. \cite{SinghSS}. 
$v_{\mB, T}$ agrees with the lower-mode velocity and shows the crossover behavior at the transition, rather than a jump, 
which is expected for a finite-size system \cite{Pilati2008, Matt2010, sf_2019, SinghJJ}.

In Fig. \ref{Fig:vel} we now compare the DSF results with the sound velocities determined using density probes in the experiments and in our simulations. 
Overall, the measured higher-mode velocity agrees with the DSF higher-mode velocity. 
However, the measured lower velocity is higher than the DSF lower velocity at $T/T_0 < 1$, except for the measurements at $T/T_0 \geq 1$, 
which show agreement. 
We note that the measured $T_c/T_0 = 1$ is determined based on the disappearance of lower sound peak \cite{Hadzibabic2021}.
We also present the simulation velocities obtained by imitating the experimental protocol described in Sec. \ref{sec:excitation}. We used the driving strength $V_0/\mu=1$ in line with the used values in the experiments between $V_0/\mu=0.47 - 1$ \cite{Hadzibabic2021}.
The simulated higher-mode velocity agrees with the measured higher-mode velocity and shows deviations from the DSF higher-mode velocity at low temperatures. 
On the other hand, the simulated lower velocity is smaller than the measured lower velocity at $T/T_0 < 0.75$ and agrees with the measurements at $T/T_0 > 0.75$.
In Fig. \ref{Fig:vel} we also show the two sound velocities determined via a step-pulse perturbation in Sec. \ref{sec:step}, which agree with the results of the DSF throughout the transition for both the higher and lower velocity.

  The lower-velocity results of the DSF and step-pulse perturbation are in agreement with the Bogoliubov estimate $v_{\mB, T}$,  confirming that the sound velocity is smaller than the measurements and there is smooth crossover occurring near the transition point. 
The deviation near the transition temperature is reproduced by the simulations of the driven response, 
which yield similar velocities as in the measurements.  These simulations then systematically deviate from the measurements at intermediate and low temperatures, which seems to occur due to a variation in the used value of the driving strength in the measurements and the corresponding change in nonlinear response. Furthermore, the measurement uncertainty of the box length and the density can also affect the magnitude of the measured sound velocities \cite{Hadzibabic2021}.

\section{Conclusions}\label{sec:conclusion}
We have determined and discussed the propagation of first and second sound in homogeneous 2D Bose gases across the BKT transition using classical-field simulations for the experimental parameters of Ref. \cite{Hadzibabic2021}.
We have identified the two sound modes based on the dynamic structure factor, which are the Bogoliubov (B) and non-Bogoliubov (NB) sound modes below the transition and the diffusive and normal sound modes above the transition. 
We have excited the sound modes using the experimental method of periodic driving \cite{Hadzibabic2021} and the method of step-pulse perturbation \cite{Hoffmann2021, SinghSS}. 
We have determined the sound velocities from the dynamic structure factor (DSF), the driven response and the step-pulse excitation and compared them with the measurements of Ref. \cite{Hadzibabic2021}. 
While the sound results of the DSF and step-pulse excitation show excellent agreement, the results of the driven response show a systematic deviation, compared to the DSF results,  due to nonlinear response. 
If the probing strength is small, the driven response recovers the Bogoliubov mode velocity. However, for small probing strengths, the signal of the non-Bogoliubov mode vanishes. It only becomes measurable for intermediate probing strengths. At these probing strengths, the nonlinear character of the probe influences the frequencies of the measured sound modes. 
Therefore, this approach only gives an approximate value of the sound velocities, for these probing strengths.
Overall, the simulated higher-mode velocity is above the B mode velocity and displays a weak-temperature dependence across the transition, which is in agreement with the corresponding measurements of the higher mode velocity. 
On the other hand, the measured lower mode velocity is below the simulated velocities of the DSF and step-pulse excitation but agrees with the results of the simulated density response at high temperatures across the transition.

   Our results give insight into the temperature dependence of the two sound modes of dilute 2D Bose gases, and the signature of these modes in the measurement techniques of Refs. \cite{Hadzibabic2021} and \cite{Hoffmann2021}. Our results reproduce largely the measurement results of \cite{Hadzibabic2021}, while demonstrating that the strong-driving results are subject to driving-induced frequency shifts. These shifts might obscure the measurements of these frequencies. Furthermore, generating a signal for the non-Bogoliubov mode requires strong driving, making this probe technique more suitable for a qualitative investigation of the modes, in contrast to the step-pulse technique of \cite{Hoffmann2021}.   
For increasing interactions or densities, these modes undergo hybridization and crossover to the strong-coupling regime occurs \cite{SinghSS, SinghSF}, which warrant further experimental investigations. 
The two sound modes and their coupling can affect the dynamics, such as the propagation of deterministic vortex colliders \cite{Kwon2021}.
Our results enable the further study of these phenomena, as they provide an in-depth insight into the key probing techniques of the field.

\section*{Acknowledgements} 
We thank Panagiotis Christodoulou for insightful discussions.  
This work is supported by the Deutsche Forschungsgemeinschaft (DFG) in the framework of SFB 925 – project ID 170620586 
and the excellence cluster  `Advanced Imaging of Matter’ - EXC 2056 - project ID 390715994, and
 the Cluster of Excellence ‘QuantumFrontiers’ - EXC 2123 - project ID 390837967.

\bibliography{References}

\end{document}